%% file: main_arxiv.tex
\documentclass[twocolumn]{aastex701}

\usepackage{CJKutf8}

\usepackage{graphicx,times} 
\usepackage{natbib}
\usepackage{tcolorbox}
\usepackage{physics}
\usepackage{amssymb,amsmath}
\usepackage{indentfirst}
\usepackage{threeparttable}

\usepackage{graphicx} 
\usepackage{physics}
\usepackage{ragged2e}
\usepackage{rotating}
\usepackage{amsmath}
\usepackage[version=4]{mhchem}
\usepackage{threeparttable}
\usepackage{booktabs}

\definecolor{forestgreen}{rgb}{0.13, 0.55, 0.13}

\newcommand{\E}[1]{\times 10^{#1}}

\newcommand{\lt}{\left}       
\newcommand{\rt}{\right}

\newcommand{\kms}{km\,s$^{-1}$}
\newcommand{\Tk}{T_{\rm k}}
\newcommand{\Tex}{T_{\rm ex}}
\newcommand{\nHH}{n({\rm H}_{2})} 
\newcommand{\NSiO}{N({\rm SiO})} 
\newcommand{\VLSR}{$V_{\rm LSR}$}
   
\newcommand{\tto}{2--1}
\newcommand{\ftf}{5--4}
\newcommand{\ets}{8--7}

\newcommand{\sztso}{$7_0$--$6_1 A^+$}
\newcommand{\Jsztso}{$J$=\sztso}

\newcommand{\Jto}{$J$=\tto}
\newcommand{\Jff}{$J$=\ftf}
\newcommand{\Jes}{$J$=\ets}

\newcommand{\HtCO}{H$_2$CO}
\newcommand{\CHthOH}{CH$_3$OH}

\newcommand{\hii}{H\textsc{ii}} 
\newcommand{\jypbm}{\mbox{Jy\,beam$^{-1}$}}
\newcommand{\mjypbm}{\mbox{mJy\,beam$^{-1}$}}

\newcommand{\target}{Sgr~C-F\,\#2}

\newcommand{\msun}{$M_{\odot}$}
\newcommand{\cc}{\mbox{cm$^{-3}$}}
\newcommand{\sqc}{\mbox{cm$^{-2}$}}
\renewcommand{\arcsec}{$^{\prime\prime}$}

\usepackage{hyperref}

\begin{document}
\begin{CJK}{UTF8}{gbsn}

\title{A Collimated Protostellar Outflow Likely Driven by Jet and Reshaped by Environment in the Sagittarius C Cloud of the Central Molecular Zone}

\correspondingauthor{Sen-di Bo, Xing Lu}
\email{bosendi26@mails.ucas.ac.cn, xinglu@shao.ac.cn}

\input{authors}

\begin{abstract}
  The Central Molecular Zone (CMZ) at the center of the Galaxy provides a nearby laboratory for studying star formation under extreme physical and chemical conditions. 
  Recent ALMA (Atacama Large Millimeter/Submillimeter Array) observations have identified 43 candidate protostellar outflows in three CMZ clouds. One of these, the \target{} outflow in Sagittarius C (Sgr C), has a highly collimated lobe extending approximately 0.8~pc in projection. Here we present a joint analysis of ALMA observations of SiO \Jes, \ftf,  and \tto. 
  We use local thermodynamic equilibrium (LTE) and non-LTE methods to constrain the SiO column densities, H$_2$ number density, and gas temperature. The low SiO rotational temperatures and higher kinetic temperatures inferred from non-LTE methods suggest subthermal excitation. 

  The elongated SiO morphology, shock-excited gas, position-velocity structure, and the alignment of the red-shifted molecular lobe with JWST infrared knots altogether favor jet-dominated entrainment, whereas a broader disk-wind component close to the source cannot be excluded. The outflow shows asymmetries in its morphology and physical properties, which can be attributed to dense gas on the path of its lobe impeding its propagation and feedback from a nearby \hii{} region imposing different pressure on its two sides.
\end{abstract}


\section{Introduction}
\label{sec:Introduction}

Protostellar outflows are an essential component in star formation. Stars form through the gravitational collapse of over-densities in molecular clouds \citep{Shu1987,McKeeOstriker2007}. As collapse continues, conservation of angular momentum naturally leads to the formation of a rotationally supported circumstellar disk around the protostar. Without efficient angular-momentum transport, centrifugal support would inhibit further accretion. Magnetized disk winds, jets, and the molecular outflows that they entrain provide a way for removing angular momentum and mass from the star-disk system, enabling gas in the disk to migrate inward and accrete onto the protostar \citep{BlandfordPayne1982,Shu1994,KoniglPudritz2000,Pudritz2007,Frank2014,Bally2016}. The outflows also constitute an important component of protostellar feedback. By injecting momentum and kinetic energy into the surrounding gas, they can drive turbulence, disperse gas, and help regulate the local star formation efficiency and the initial mass function\citep{MatznerMcKee2000,NakamuraLi2007,Frank2014,Bally2016,guszejnov2021}.

In high-mass ($M_\ast \gtrsim$ 8~\msun{}) star formation, outflows are especially important because accretion proceeds in the presence of strong radiative feedback \citep{ZinneckerYorke2007,Tan2014}. Radiation from the growing massive protostar and its accretion flow can exert substantial pressure on the dusty infalling envelope \citep{Keto2006}. Disk-mediated accretion and bipolar outflow cavities help make the radiation field anisotropic, allowing radiation to escape preferentially along the polar directions. This reduces radiative feedback against equatorial accretion and helps sustain the growth of massive protostars \citep{YorkeSonnhalter2002,Krumholz2009,Kuiper2015}. Observed correlations between outflow properties and source luminosity over several orders of magnitude further suggest that massive protostellar outflows are connected to the same accretion-ejection processes observed in low-mass systems, likely involving magneto-centrifugal disk-wind or jet-launching mechanisms, but operating at higher mass-loss rates, momenta, and kinetic energies, and often with more complex or wider-angle morphologies \citep{Beuther2002,Richer2000,Maud2015}.

The Central Molecular Zone (CMZ), the inner $\sim$200--500~pc of the Galaxy, is the nearest galactic center environment in which individual sites of star formation can be resolved. Compared to typical molecular clouds in the Galactic disk, CMZ clouds are characterized by elevated gas temperatures ($T_{\rm k}\sim 50$--$200$~K) and H$_2$ number densities ($n_{\rm H_2}\gtrsim 10^{4}$~\cc{}), highly turbulent motions (Mach number $\sim$ 10--20), strong magnetic fields ($\sim$mG), and enhanced cosmic-ray ionization rates \citep[10$^{-15}$--10$^{-13}$~s$^{-1}$; see][and references therein]{Henshaw2023}. Although the CMZ contains $\gtrsim 10^{7}$~\msun{} of dense molecular gas, its star formation efficiency (star-formation rate per unit dense gas mass) is lower by at least an order of magnitude than expected from empirical star-formation relations calibrated in the Galactic disk \citep{Longmore2013,Barnes2017,Lu2019a,Lu2019b}.

Identifying individual protostellar outflows in the CMZ is observationally challenging. The low star formation efficiency suggests fewer outflows per unit gas mass than in typical Galactic disk clouds. Severe optical and near-infrared extinction, source crowding, complex line-of-sight gas structures, and widespread shocks unrelated to protostellar feedback further complicate their identification \citep{Jones2012,Yang2025,Battersby2026}. Therefore, high-resolution interferometric observations are crucial. Previous studies have identified protostellar outflows in the Sagittarius B2 (Sgr B2) cloud, one of the most active star forming regions in the Galaxy \citep{Qin2008,Higuchi2015,Meng2022,Busch2024,Budaiev2025,Budaiev2026}. 

More recently, ALMA observations at $\sim$2000~au resolution have revealed outflows in CMZ clouds outside of Sgr~B2. \citet{Lu2021} used multiple shock tracers, including SiO, SO, \CHthOH{}, \HtCO{}, HC$_3$N, and HNCO, to identify 43 bipolar or monopolar outflows in three massive CMZ clouds. \citet{Walker2021} found multiple outflows traced by SiO emission in the seemingly quiescent G0.253$+$0.016 cloud. In addition, near-infrared observations with the James Webb Space Telescope (JWST) have revealed shocked molecular gas in protostellar outflows in the Sagittarius C (Sgr~C) cloud \citep{Crowe2025}. These observations establish that outflow activity occurs in several CMZ star forming regions.

\begin{figure*}[t]
    \centering
    \includegraphics[width=0.98\linewidth]{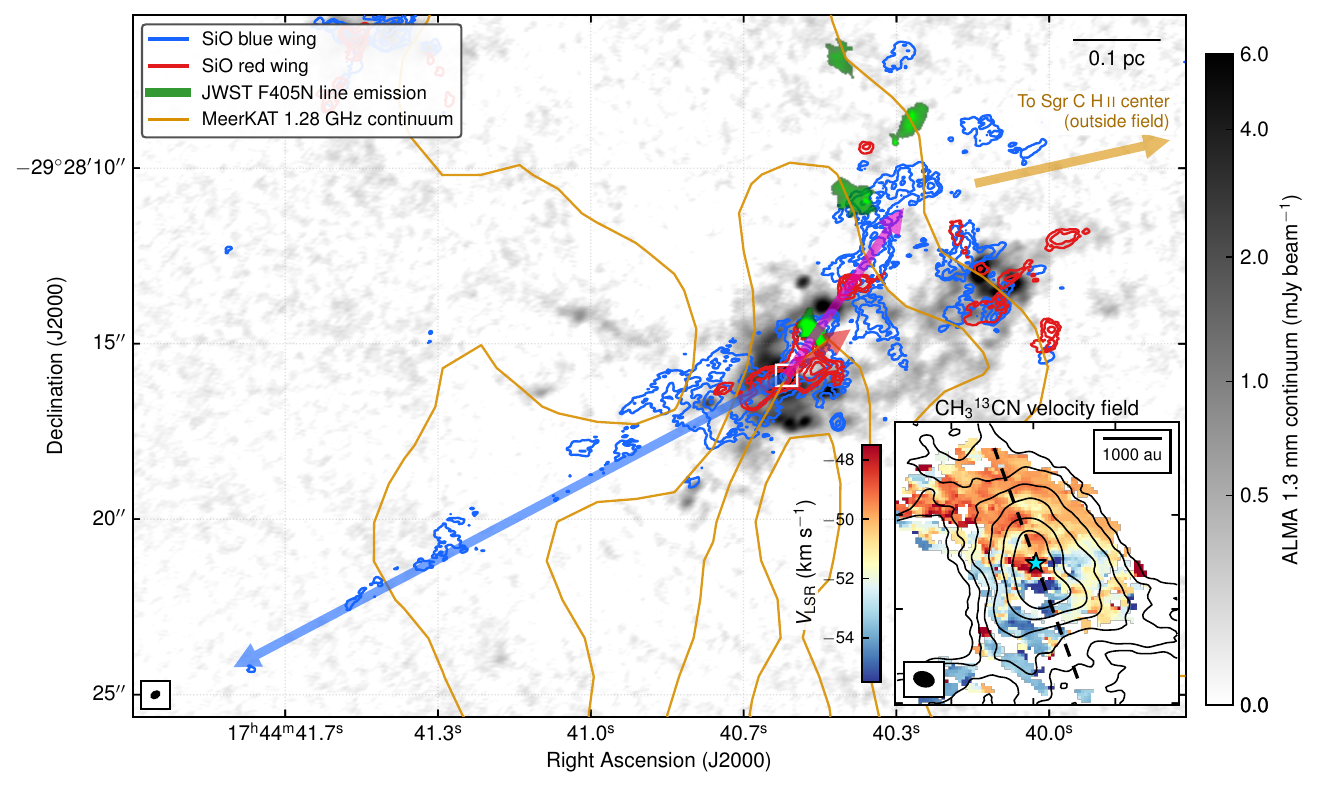}
    \caption{Multi-wavelength view of the \target{} outflow. The main panel shows the ALMA 1.3~mm continuum emission in grayscale. Blue and red contours show blue-shifted and red-shifted SiO \Jff{} integrated intensities at [0.03, 0.06, 0.12, 0.18]~\jypbm{}\,\kms{} \citep{Lu2020,Lu2021}. The corresponding arrows indicate the projected directions of the two SiO lobes. Green patches show the JWST/NIRCam F405N emission, with compact spurious sources suppressed for display, and the magenta arrow marks the candidate jet \citep{Crowe2025}. Orange contours show the MeerKAT 1.28~GHz continuum at intensities of [4, 5, 6, 7]~\mjypbm{} at $4''$ resolution \citep{Heywood2022}, outlining the extended radio emission associated with the adjacent Sgr~C \hii{} region. The orange arrow points toward the center of the \hii{} region defined by \citet{DeBuizer2025}, outside the field to the northwest. The white square marks the region shown in the lower-right inset. The inset presents the CH$_3^{13}$CN velocity field in color, overlaid with high-resolution 1.3~mm continuum contours in black at 0.085$\times$2$^n$~\mjypbm{}, $n=0$--6 \citep{ZhangS2025}. The cyan star marks the protostar, and the black dashed line shows the orientation of the fitted velocity gradient. Synthesized beams are shown in the lower-left corners. Scale bars correspond to 0.1~pc in the main panel and 1000~au in the inset.}
    \label{fig:overview}
\end{figure*}

Despite this growing sample, spatially resolved studies on the excitation of outflows in the CMZ remain limited, impeding comparisons with their counterparts in the Galactic disk. To address this issue, we select a well-resolved outflow in Sgr~C (see \autoref{fig:overview}; \target{} hereafter) for a multi-wavelength study. In particular, we combine three SiO transitions from independent ALMA projects to constrain the gas excitation and discuss the origin and evolution of the outflow. This is the first paper of the Coordinated Homogenized Observations from Reprocessed Uncoordinated Surveys of the CMZ (CHORUS) series, in which we collect data from complementary observations to study outflows, hot cores, and filaments in the CMZ. Subsequent papers in the series will extend this analysis to a large sample of outflows.

This paper is organized as follows. In \autoref{sec:Observations and Data Reduction}, we describe the observations and data reduction procedures. In \autoref{sec:Results}, we present the morphology and kinematics of the outflow, and derive densities and temperatures. In \autoref{sec:Disscussion}, we discuss the asymmetries and driving mechanisms of the outflow, and implications for star formation in the CMZ. Finally we summarize our findings in \autoref{sec:conclusion}. Throughout this paper we adopt a distance to Sgr~C of 8.277~kpc, consistent with the Galactic-center distance measured by \citet{Gravity2022}.

\section{Observations and Data Reduction}
\label{sec:Observations and Data Reduction}

\subsection{The Target: \target{} Outflow}\label{subsec:obs_target}

Early evidence for protostellar outflows in Sgr~C came from collisionally excited H$_2$O masers \citep{Chambers2014,Titmarsh2016,Lu2019a} and extended green objects \citep[][]{Chen2013,Kendrew2013,Chambers2014}. \citet{Kendrew2013} detected compact H$_2$ and Br\,$\gamma$ emission in infrared slit spectra, which support the existence of outflows, although the exact morphology and orientation of the outflows were unconstrained. The Submillimeter Array (SMA) observations did not resvolve outflows, probably because of limited angular resolution and sensitivity \citep{Lu2019a}. Finally, the outflows were unamnbigously detected by using ALMA SiO $J$=5--4 line emission in \citet{Lu2021}. A sample of 18 outflows, including \target{}, were identified within the observed portion of the cloud.

We selected \target{} because its $\sim$0.8-pc projected extent permits a resolved study along the major axis (\autoref{fig:overview}). The velocity of SiO $J$=5--4 does not show a clear gradient along the major axis, consistent with an axis close to the plane of the sky (see Figure~23 of \citealt{Lu2021}). It is associated with a hot core that hosts a class \textsc{ii} \CHthOH{} maser and an ultra-compact H\textsc{ii} region, both suggesting high-mass star formation \citep{Caswell2010,Lu2019a,Lu2019b}. The core itself shows a velocity gradient aproximately perpendicular to \target{} in CH$_3$$^{13}$CN emission (\autoref{fig:overview} inset), indicating rotation. All these features point to a potential high-mass protostellar disk-outflow system in this source.

\subsection{ALMA Band~1 Observations}\label{subsec:obs_b1}

We conducted new Band~1 observations towards Sgr~C in 2025 May and September with the C-6 and C-9 configurations, respectively, under the project 2024.1.00213.S. The spectral setup placed a series of 58.59~MHz windows, sampled at a channel with of 61~kHz and a spectral resolution of 122~kHz ($\sim$0.84~\kms{}), on several spectral lines including 
the 44.0694~GHz \CHthOH{} $7_0$--$6_1\,A^+$ transition. A 1.875-GHz window was included for continuum and calibration. The \CHthOH{} transition is a collisionally pumped Class~I maser and therefore traces shocked gas \citep{Menten1991}. We calibrated the delivered data with the ALMA pipeline in CASA \citep{CASA2022}.

\subsection{ALMA Band~3 Observations}\label{subsec:obs_b3}

SiO~\Jto\ at 86.8470~GHz was extracted from the DUET observations (project code: 2018.1.01420.S; \citealt{Xu2025}), which combine the C43-4 configuration observed in 2018 November and the C43-7 configuration observed in 2019 August and 2021 July--August. The baseline ranges were 15~m--1.4~km and 38~m--5.2~km, respectively. The SiO line falls in a 0.9375~GHz spectral window centered at 86.71~GHz with a native channel width of 0.244~MHz and a spectral resolution of 0.488~MHz ($\sim$1.67~\kms{}). Each epoch was pipeline-calibrated using its associated CASA version. After line-free channels were identified, data from the two configurations were concatenated and the continuum was subtracted.  

\subsection{ALMA Band~6 Observations}\label{subsec:obs_b6}

The Band~6 data were obtained in the C40-3 and C40-5 configurations in 2017 April and July (project code: 2016.1.00243.S). The correlator covered 217--221 and 231--235~GHz with a spectral resolution of 1.129~MHz, corresponding to $\sim$1.5~\kms{}, and a channel width of 1.35~\kms{}. 
The SiO~\Jff{} line at 217.1050~GHz was employed in this study.

We used the standard ALMA pipeline to calibrate the execution blocks separately. Then the configurations were combined and imaged in CASA. We identified line-free visibility channels and subtracted the continuum. The data have been presented in \citet{Lu2020,Lu2021}, \citet{Yang2025}, and \citet{Zhang2025}.

\subsection{ALMA Band~7 Observations}\label{subsec:obs_b7}

We used the SiO~\Jes\ line at a rest frequency of 347.3306~GHz from a Cycle~8 program (project code: 2021.1.00286.S). The observations combined the C-5 configuration (15~m--1.3~km baselines; conducted in 2022 June) and the C-2 configuration (15--455~m baselines; conducted in 2022 April). Four 1.875~GHz spectral windows between 335 and 349~GHz provided a total nominal bandwidth of 7.5~GHz and included the SiO line. The channel width is 976~kHz, corresponding to $\sim$0.84~\kms{} at the observed frequency.

The data were calibrated with the standard ALMA pipeline followed by the polarization calibration script supplied by ALMA staff (see \citealt{Liu2026} for details). However, for the spectral line analysis here, we only used the total intensity products.

\subsection{Imaging of the Spectral Lines}\label{subsec:obs_further}

To make the SiO line measurements more comparable, we restricted the Bands 3, 6, and 7 datasets to their common $uv$ distance interval, 14--1502~k$\lambda$, and imaged the visibilities with the CASA \texttt{tclean} task. Matching the $uv$ interval reduces differences in spatial filtering, but does not make the two-dimensional $uv$ sampling identical.

We used Briggs weighting with a robust number of 0.5, a pixel size of 0\farcs{04}, and the multiscale deconvolver with scales of [0, 10, 30, 90] times the pixel size. The image size was chosen to cover the relevant field.  Following \citet{Lu2021}, We ran the \texttt{tclean} task twice, first with auto-masking, and then with a broad mask bounded by the primary beam response levels of 0.3 for Band 3, 0.25 for Band 6, and 0.15 for Band 7. The images were CLEANed down to about 3$\sigma$. Within each cube, channel-dependent beam variations due to slightly different $uv$ samplings were removed by convolution to the corresponding smallest common beam determined using the \texttt{common\_beam} method with $tolerance = 0.1$ in the \texttt{radio-beam} package.

The resulting synthesized beams of the SiO cubes, from \Jto{} to \Jes{}, are $0\farcs{46}\times0\farcs{31}$ (PA $=101\fdg{3}$), $0\farcs{36}\times0\farcs{27}$ (PA $=122\fdg{6}$), and $0\farcs{31}\times0\farcs{20}$ (PA $=79\fdg{0}$), respectively. Position angles (PAs) are measured east of north. The residual beam differences reflect the non-identical two-dimensional $uv$ sampling.

For the excitation analysis, the three SiO cubes were further convolved to a nominal common beam of $0\farcs{46}\times0\farcs{31}$ with a position angle of 101\fdg{3} and a channel width of 1.35~\kms{}. The beam is the smallest common beam size of the four cubes, while the channel width is the broadest among the cubes.
The root mean squared (RMS) noise levels of the resampled SiO~\Jes, \Jff, \Jto\ cubes are 0.26, 0.38, and 0.77~K per channel, respectively. 

For the compact \CHthOH~\Jsztso\ masers, we retained the full $uv$ range since recovery of diffuse line emission is not required to locate compact sources. The angular resolution is $0\farcs{30}\times0\farcs{21}$ while the channel width is 122~kHz, corresponding to 2500~AU $\times$ 1700~AU and 0.83~\kms, respectively.

\subsection{Archival Infrared and Radio Data}\label{obs:archival}

For the multiwavelength comparison, we used the James Webb Space Telescope (JWST) Near Infrared Camera (NIRCam) observations of Sgr C presented by \citet{Crowe2025}. In particular, the continuum-subtracted F470N image contains the H$_2$ 0--0 S(9) line, F212N contains the H$_2$ 1--0 S(1) line, and F405N contains Br\,$\alpha$, all of which reveal a likely lobe of the jet associated with the \target{} outflow. We also inspected the associated continuum-filter images to distinguish compact line-emission knots from emission from the massive protostar itself. The comparison with the ALMA SiO emission is described in \autoref{subsec:Driving Mechanism of the Outflow}.

We also obtained the 1.28~GHz MeerKAT continuum image toward the Galactic Center \citep{Heywood2022}, which delineates the spatial extent of the Sgr~C \hii{} region and provides environmental context.

\section{Results}
\label{sec:Results}

\subsection{Morphology of the Outflow}
\label{subsec:Spatial Distributions of the Outflow}

\begin{figure*}
    \centering
    \includegraphics[width=0.98\linewidth]{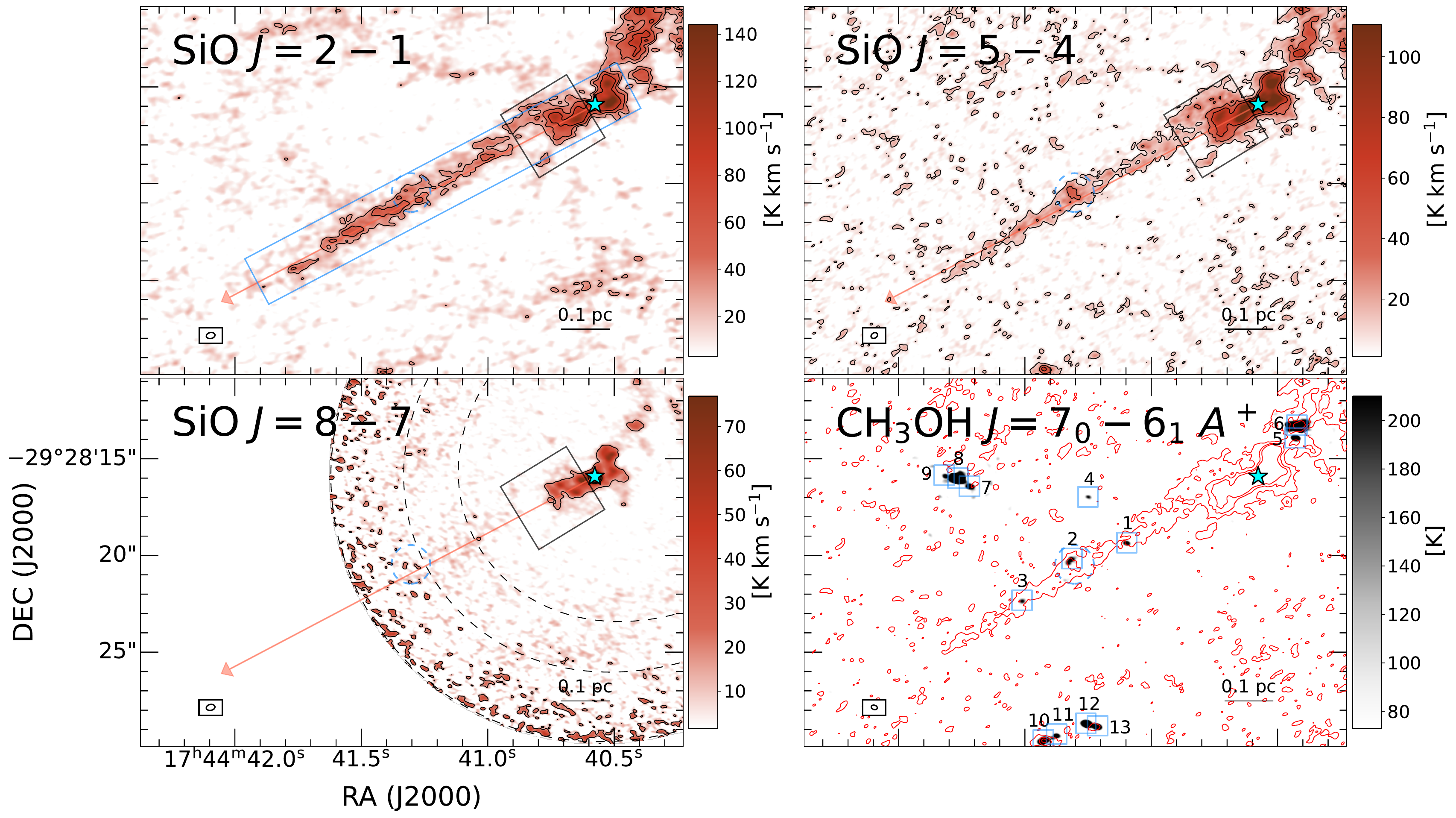}
    \caption{Integrated brightness temperature over \VLSR{} $=-69$ to $-42$~\kms{}, of SiO~\Jto, \Jff, and \Jes{}, respectively, which is overlaid with their 1$\times$, 3$\times$ and 5$\times$ RMS noise levels. The bottom-right panel shows the peak brightness temperature of the Class~I \CHthOH{}~\Jsztso{} maser transition, and the red contours outline 1$\times$, 3$\times$ and 5$\times$ RMS noise levels of the SiO~\Jff{} outflow for comparison. 
    The brightness-temperature scales of the first three panels are in unit of K\,\kms{}, while that of the last panel is is unit of K. The cyan stars mark the protostellar position. The arrows in the first three panels indicate the projected direction of the southeastern blue-shifted lobe, which are the same as the blue arrow in \autoref{fig:overview}. The blue rectangle in the top-left panel indicates the region where we extract the spectra along the axis in \autoref{fig:PV diagrams}. The black box in the first three panels marks the region of the wide-angle component in the southeastern lobe, whereas the blue dashed circle marks the region where SiO~\Jff{} is locally enhanced. The black dashed contours in the bottom-left panel indicate the 0.1, 0.3, and 0.5 primary beam response levels for SiO~\Jes. The blue squares in the bottom-right panel mark the identified masers. 
    Synthesized beams are shown in the lower-left corners and scale bars in the lower-right correspond to 0.1~pc.}
    \label{fig:Moment0 maps}
\end{figure*}

The \target{} outflow is detected in the three SiO transitions 
(\autoref{fig:Moment0 maps}). We adopted the hot core systemic velocity $v_\text{sys}=-50.2$~\kms{} from \citet{Lu2021}, which is consistent with the velocity field in \autoref{fig:overview} inset. Then we integrated the intensities of the SiO lines within local standards of rest velocities ($V_{\rm LSR}$) of $-69$ to $-42$ \kms, and displayed the maps in \autoref{fig:Moment0 maps}.

The three lines generally trace the same elongated structure. Along much of the collimated lobe, the integrated SiO brightness temperature decreases with increasing rotational quantum number, 
\Jto{} $>$ \Jff{} $>$ \Jes. 
The low-$J$ SiO lines (\Jff, \Jto) well delineate the collimated lobe profile, whereas the \Jes{} line emission mostly concentrates around the hot core and becomes weaker along the lobe. 

We observed several morphological features of this outflow:
\begin{itemize}
\item  
The southeastern outflow lobe is made up of a wide-angle component near the hot core, marked by the black box in \autoref{fig:Moment0 maps}, and a collimated, jet-like component. The angular length of this lobe is approximately 19\farcs{5}, corresponding to $\sim$0.8~pc in projection, and is about 1\arcsec{} ($\sim$8000~AU) wide over much of its length, as seen in SiO \Jto{} and \Jff{}. The field of view of the \Jes{} observations is smaller, only covering part of the lobe closer to the hot core. In all lines, the projected lobe width tends to decrease away from the hot core.
\item 
The integrated brightness of SiO~\Jff\ transition is locally enhanced within the blue dashed circle in \autoref{fig:Moment0 maps}. The corrresponding \Jto{} emisssion is also enhanced compared to the inner portion of the lobe closer to the core, while the \Jes{} emission cannot be clearly seen as it is on the edge of the primary beam.
\item
Fourteen compact, bright, and narrow-line \CHthOH{} \Jsztso{} sources are detected near the outflow and are candidate class I masers. The bottom right panel in \autoref{fig:Moment0 maps} presents the peak intensity map of \CHthOH~\Jsztso\ transition where the positions of candidate maser spots are marked. The peak brightness temperatures, intrinsic full width at half maximum (FWHM) linewidths where the channel width has been subtracted in quadrature, and \VLSR{} of these candidate masers are listed in \autoref{tab:maser}. 
Among them, \#1, \#2a, \#2b, and \#3 are associated with the blue-shifted outflow lobe in the southeast, and \#5 and \#6 are associated with the red-shifted gas on the other side. In particular, \#2b and \#2b are spatially coincident with the enhanced SiO~\Jff{} emission described in the last item.
\item
The northwestern red-shifted lobe is not clearly detected. Molecular line emission can be seen; however, it is not as collimated as the other lobe. Instead, it is shorter and broader. Such lobe-to-lobe asymmetry has been found in nearby star forming regions \citep[e.g.,][]{Wassell2006,Assani2024,bouvier2025}, and will be discussed in \autoref{subsec:asym_twolobes}.
\item 
The molecular emission is distributed unevenly perpendicular to the visually adopted major axis that is illustrated by the arrow in \autoref{fig:Moment0 maps}, with more prominent SiO emission on its northeastern side. The axis position is uncertain; therefore we only consider this as a qualitative morphological asymmetry. This will be discussed further in \autoref{subsec:asym_transverse}.

\end{itemize}

\subsection{Kinematics of the Outflow}

\begin{figure*}[ht]
    \centering
    \includegraphics[width=0.98\linewidth]{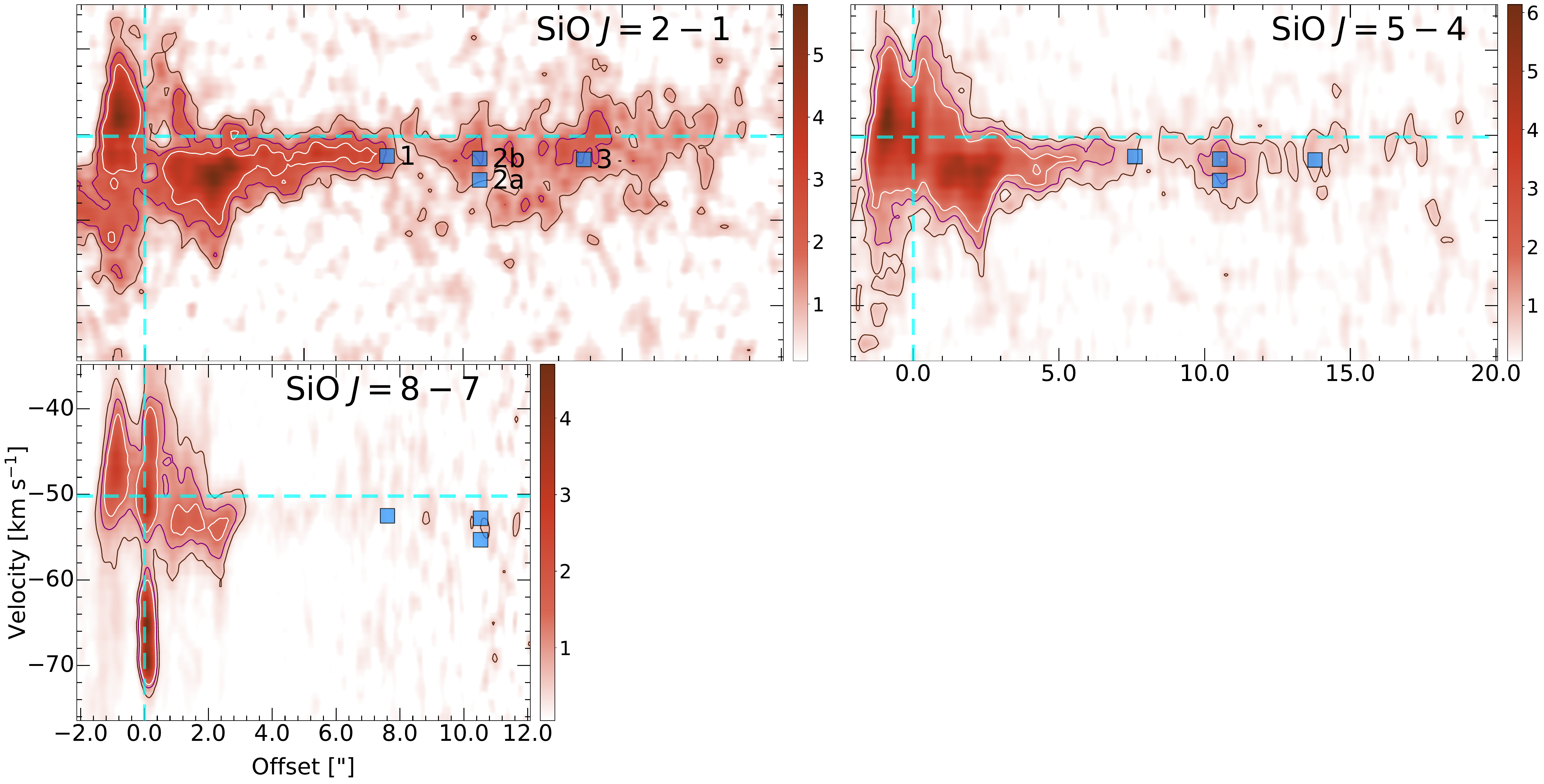}
    \caption{Position--velocity diagrams along the southeastern blue-shifted outflow lobe for the four SiO transitions shown in the region of \autoref{fig:Moment0 maps}. Color indicates the averaged brightness temperature, with 1$\times$, 2$\times$, 3$\times$ RMS noise levels as contours to emphasize the line-emission structure. The RMS values are 0.87~K for \Jto{}, 0.53~K for \Jff{}, 0.47~K for \Jes{}. 
    The brightness-temperature scales of three panels are in unit of K.
    The horizontal dashed cyan lines mark the hot core systemic velocity $\rm v_\text{sys}=-50.2$~\kms{} from \citet{Lu2021}. The vertical dashed cyan lines mark zero axial offset at the base of the lobe, while positive offset increases toward its southeastern tip. The blue squares mark maser spots in position-velocity diagrams. SiO~\Jes{} uses smaller extraction aperture in \autoref{fig:Moment0 maps}, while the other transitions use the larger aperture.}
    \label{fig:PV diagrams}
\end{figure*}

To identify the velocity components of the southeastern blue-shifted lobe, we extracted position–velocity (PV) diagrams along the projected direction of the southeastern outflow lobe within the 21\farcs{7} $\times$ 2\farcs{7} blue rectangular aperture defined in \autoref{fig:Moment0 maps}.
\autoref{fig:PV diagrams} shows the PV diagrams of the collimated outflow for the three SiO lines.
Near the hot core, strong and broad emission is seen over \VLSR{} $=-76$ to $-34$ \kms with complex variations, indicating multiple velocity components. 
At offsets of $\sim$5\arcsec{}--8\arcsec{}, the emission narrows to $-55$ to $-50$~\kms, and its centroid velocity approaches $v_\text{sys}$. Beyond the offset of 10\arcsec{}, which corresponds to the location of the enhanced SiO \Jff{} emission mentioned in \autoref{subsec:Spatial Distributions of the Outflow}, the SiO emission broadens again. 
The four Class~I \CHthOH{} masers have velocities around \VLSR $= -52$ \kms, matching those of SiO in the outflow.

\subsection{Physical Properties of the Outflow}
\label{subsec:Physical Properties of the Outflow}

\begin{figure*}[ht]
    \centering
    \includegraphics[width=0.98\linewidth]{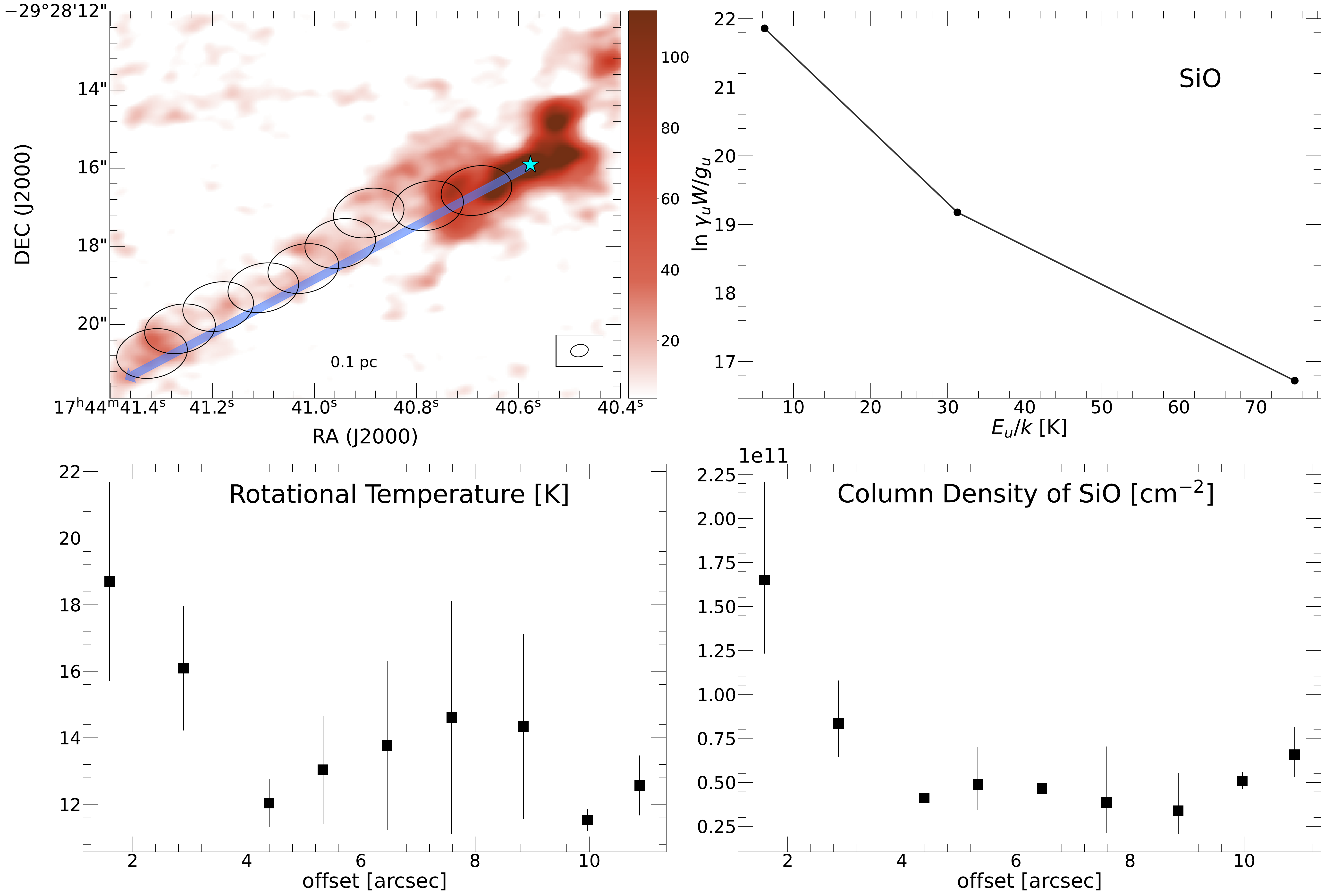}
    \caption{
    The top left panel shows the integrated intensity map of smoothed SiO~\Jff{}. The brightness-temperature scale is in unit of K\,\kms{}.
    The cyan star marks the protostellar position and the blue arrow indicates the projected direction of the southeastern blue-shifted lobe. 1\farcs{84} $\times$ 1\farcs{24} black ellipses mark regions where we extracted spectra for LTE and non-LTE analysis. The synthesized beam is shown in the lower-right corner and scale bar in the lower center corresponds to 0.1 pc. 
    The top right panel shows the representative rotation diagram in one aperture. The horizontal axis indicates the energy levels of transition while the vertical axis indicates the integrated intensities.
    The bottom two panels show the variations of SiO excitation temperature and column density estimated from LTE rotation diagram method. The offsets are defined by the separation between the centers of ellipse apertures and the protostellar position and increase to the southeast.}
    \label{fig:Rotation map}
\end{figure*}

\subsubsection{Local Thermodynamic Equilibrium Method}
\label{subsubsec:Local Thermodynamic Equilibrium Method}

Our SiO observations cover three rotational transitions from the low-energy \Jto\ to the high-energy \Jes\ line. 
Thus the rotation diagram method can be employed to constrain the physical properties of the SiO gas, by assuming local thermodynamic equilibrium (LTE) and optically thin emission \citep{Goldsmith1999}.

According to the rotation diagram method, the integrated intensity $W$ follows
\begin{equation}\label{equ:rd}
    \mathrm{ln} \frac{\gamma_u W}{g_u} = \mathrm{ln} N - \mathrm{ln} Z - \frac{E_u}{k_BT_\text{rot}},
\end{equation}
in which $\gamma_u$ is a constant indicating radiation ability, $g_u$ is the statistical weight of the upper level, $N$ is the molecular column density, $Z$ is the molecular partition function, $E_u$ is the upper energy level, $k_B$ is the Boltzmann constant and $T_\text{rot}$ is the excitation temperature. We adopted the method of \citet{Goldsmith1999} to calculate the molecular partition function and degeneracy for linear SiO molecules and retrieved the energy levels of each transition from the LAMDA database \citep{LAMDA2005} to construct the rotation diagram. 

We extracted mean SiO spectra from the 1\farcs{84} $\times$ 1\farcs{24} ellipse apertures in the top left panel of \autoref{fig:Rotation map} for higher signal-to-noise ratios ($>10$). These spectra were used for further LTE and non-LTE analyses. By fitting the integrated intensities against the upper energy levels of multiple transitions, we estimated the rotational temperature and the molecular column density in the apertures. For most of the apertures, the rotation diagrams are curved rather than linear (see the example in the top right panel of \autoref{fig:Rotation map}). 
This suggests the excitation in the outflow deviates from LTE, and the optical depth is non-negligible. We nevertheless retain linear fits in the rotation diagrams as an LTE reference. The resulting rotational temperatures and column densities of SiO are presented in the bottom panels of \autoref{fig:Rotation map}. The SiO rotational temperatures $T_\text{rot}$ are on the order of magnitude of 10~K while the SiO column densities $\NSiO$ are on the order of magnitude of $10^{11}$~\sqc{} in the southeastern lobe.

\begin{table}[t]
  \centering
  \caption{Candidate Class I \CHthOH~\Jsztso{} masers.
  Peak temperature of Class I \CHthOH{} \Jsztso{} transition, intrinsic FWHM, and \VLSR{} are listed. The FWHMs of maser \#7, \#9 and \#13 are reported to be narrrower than one channel width because their line profiles are not resolved.}
  \label{tab:maser}
  \begin{tabular}{ l c c c}
    \toprule
    Maser & {$T_{\rm peak}$ (K)} & {FWHM (km\,s$^{-1}$)} & {\VLSR{} (km\,s$^{-1}$)}\\
    \midrule
    \#1  & 215  & 2.33 & $-$52.5\\
    \#2a & 188  & 2.07 & $-$55.3\\
    \#2b & 158  & 1.50 & $-$52.8\\
    \#3  & 128  & 2.52 & $-$52.9\\
    \#4  & 131  & 0.69 & $-$55.2\\
    \#5  & 518  & 0.49 & $-$52.7\\
    \#6  & 2740 & 0.42 & $-$54.0\\
    \#7  & 341  & $\le$0.83 & $-$54.9\\
    \#8  & 7790 & 0.67 & $-$55.2\\
    \#9  & 186  & $\le$0.83 & $-$56.4\\
    \#10 & 493  & 1.95 & $-$62.8\\
    \#11 & 229  & 0.82 & $-$61.0\\
    \#12 & 1060 & 0.46 & $-$59.3\\
    \#13 & 645  & $\le$0.83 & $-$59.9\\
    
    \bottomrule
  \end{tabular}
\end{table}

We also fitted Gaussian line profiles with an LTE forward-modeling method \citep{Mangum2015}. By assuming finite optical depths in different transitions, we estimated the excitation temperature and molecular column density. According to \citet{Mangum2015}, the column density and excitation temperature can be expressed as
\begin{equation}\label{equ:ff N}
    N=\frac{3hZ\tau_v \Delta v}{8\pi^3g_u|\mu_{lu}|^2}{\rm exp}\left( \frac{E_u}{T_{ex}}\right)\left[{\rm exp}\left(\frac{hv}{k_BT_{ex}}\right)-1\right]^{-1},
\end{equation}
and
\begin{equation}\label{equ:ff Tex}
    T_{\rm R}=\left[J_v\left(\Tex\right)-J_v\left(T_{bg}\right)\right]\left[1-{\rm e}^{-\tau_v}\right],
\end{equation}
where $h$ is the Planck constant, $\tau_v$ is the optical depth at frequency $v$, $\Delta v$ is the FWHM of the transition, $|\mu_{lu}|^2$ is the dipole matrix element, and the Rayleigh-Jeans equivalent temperature $J_v\left(T\right)=\frac{hv/k_B}{{\rm exp}\left(\frac{hv}{k_BT}\right)-1}$. In the LTE assumption, $\tau_v=\frac{hc^2N_{jk}A_{jk}}{8\pi vk_B \Tex} \phi_{jk}$, in which $c$ is the speed of light, $N_{jk}$ is the column density, $A_{jk}$ is the spontaneous emission factor from $j$ level to $k$ level, and $\phi_{jk}$ is the profile function of the transition. We followed \citet{Goldsmith1999} to evaluate the partition function and degeneracy for linear SiO molecules. The molecular dipole moments and energy levels of the transitions were taken from the LAMDA database \citep{LAMDA2005}. In each ellipse aperture, we fitted multiple SiO spectral lines using Gaussian profiles before estimating the optical depth, the excitation temperature and the column density.

\begin{figure*}[ht]
    \centering
    \includegraphics[width=0.98\linewidth]{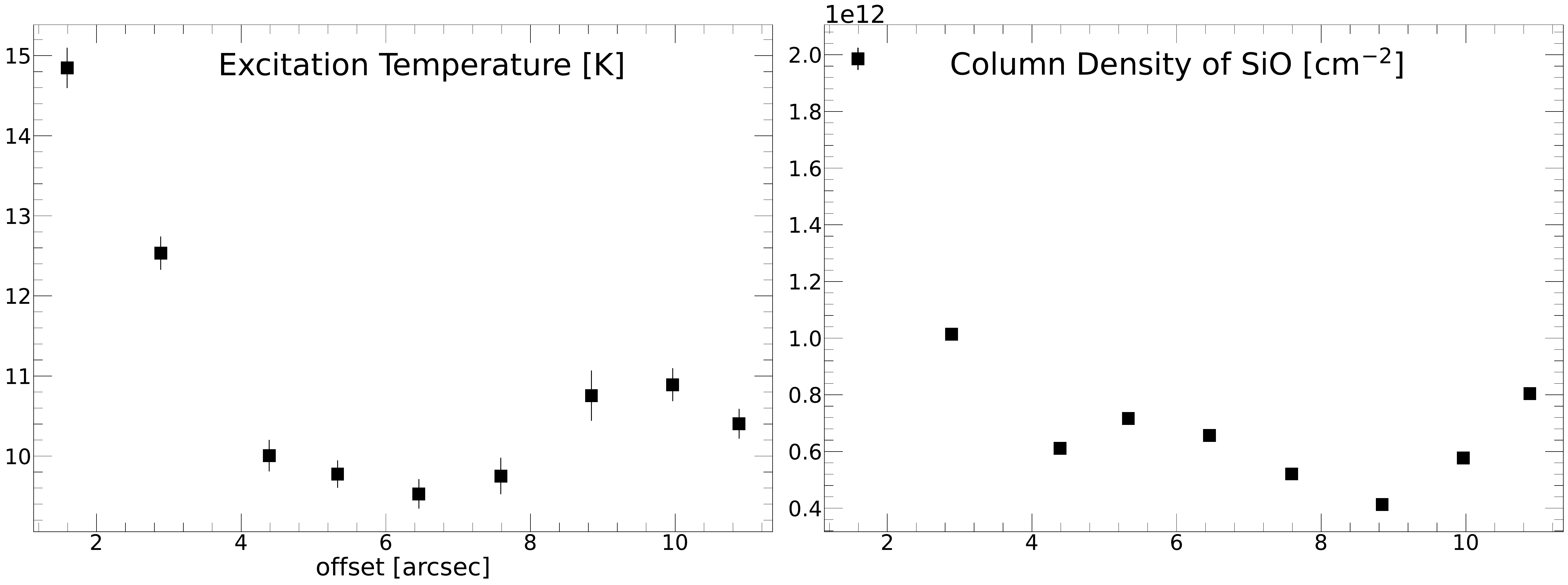}
    \caption{The variations of forward-modeling constrained excitation temperature and column density along the blue-shifted lobe. The offsets are defined by the separation between the centers of ellipse apertures and the protostellar position and increase to the southeast.}
    \label{fig:mangum column density}
\end{figure*}

The constrained excitation temperatures and column densities are shown in \autoref{fig:mangum column density}. The excitation temperatures derived are $\sim 10$ K, while the SiO column densities are on the order of magnitude of $10^{11}$--$10^{12}$~\sqc{}.

\subsubsection{Non-Local Thermodynamic Equilibrium Method}
\label{subsubsec:Non-Local Thermodynamic Equilibrium Method}
Given the non-LTE conditions indicated by the rotation diagrams, we used RADEX, which solves the molecular statistical equilibrium equations, including radiative and collisional processes, for a homogeneous isothermal component using an escape-probability treatment of radiative transfer \citep{VanderTak2007}. We further combined the Monte Carlo Markov Chain (MCMC), in particular, \texttt{emcee} \citep{emcee2013} and RADEX, to reduce running time and avoid converging to a local minimum, instead reaching the global minimum.

We used the spectra in the aperture shown in the top left panel of \autoref{fig:Rotation map}, and further extracted mean SiO spectra in the 8\farcs{2} $\times$ 0\farcs{4} apertures marked by the ellipses in the left panel of \autoref{fig:transverse_T}, of which the signal-to-noise ratios are $>$10, to obtain the distribution of physical properties transversing the lobe. We performed Gaussian fits to the \VLSR\ $\sim -52$ \kms\ component of the SiO~\Jto{}, \Jff{}, and \Jes{} lines to obtain their linewidths. Their linewidths are similar, so we adopted the linewidth of SiO~\Jff{} as a canonical value. 

By setting the logarithms of outflow kinetic temperature $\Tk$, SiO column density $\NSiO$, and $\rm H_2$ volume density $\nHH$ as fitting parameters of \texttt{emcee}, we compared the observed integrated intensities between $-$62 \kms{} and $-$42 \kms{} in addition to the intensity ratios of adjacent transitions against RADEX predictions under different physical conditions, and calculated $\chi^2$. We used the high-temperature SiO collision data from the Leiden Atomic and Molecular Database (LAMDA) and assumed that SiO collides only with H$_2$. 
The large velocity gradient (LVG) approximation was adopted. The priors of $\Tk$, $\NSiO$, $\nHH$ are 20--$10^{3.7}$~K, $10^{11}$--$10^{18}$~\sqc{}, $200$--$10^{8}$~\cc{} while they are spaced logarithmically. We set initial inputs randomly and logarithmically in the range of priors. Each MCMC run obtains 10,000 samples, of which the first half were discarded as burn-in. We used the results only as a rough constraint range, rather than taking the maximum-probability solution as the best-fit estimate. 

\begin{figure*}[t]
    \centering
    \includegraphics[width=0.98\linewidth]{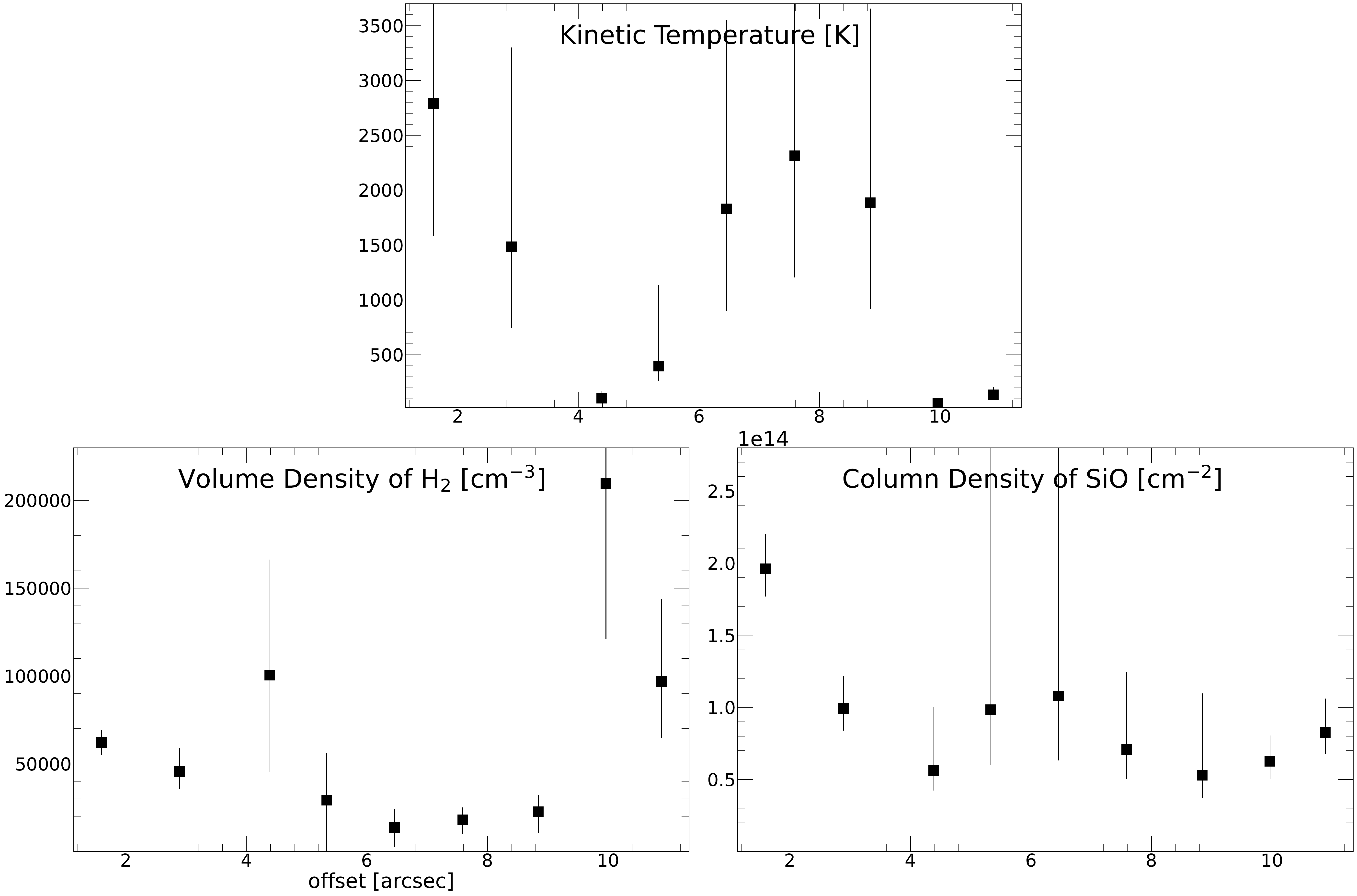}
    \caption{The variations of physical properties constrained from RADEX models along the blue-shifted lobe.
    The offsets are determined by the separation between the centers of ellipse apertures and the protostellar position and increase to the southeast.}
    \label{fig:RADEX results}
\end{figure*}

\begin{figure*}[t]
    \centering
    \includegraphics[width=0.98\linewidth]{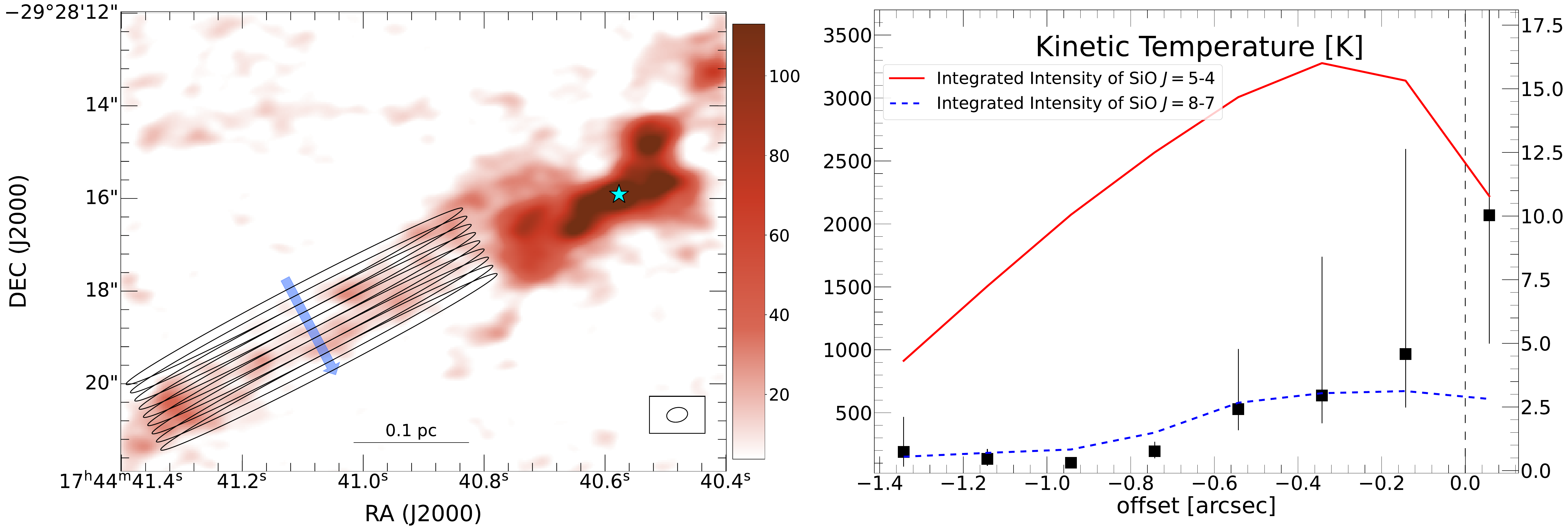}
    \caption{The left panel shows the integrated intensity map of smoothed SiO~\Jff{}. The cyan star marks the protostellar position and the blue arrow indicates the direction perpendicular to the southeastern blue-shifted lobe. 8\farcs{2} $\times$ 0\farcs{4} black ellipses mark regions where we extract spectrums to fit RADEX models. The synthesized beam is shown in the lower-right corner and scale bar in the lower center corresponds to 0.1~pc.
    The right panel shows the variations of kinetic temperatures derived from RADEX models transversing the blue-shifted lobe as well as the integrated intensities of SiO~\Jff{}, \Jes{}.
    The zero point of offset is determined on the major axis of the blue-shifted outflow. The offset is defined by the separation between the centers of ellipse and the zero point and increases to the Sgr~C \hii{} region.}
    \label{fig:transverse_T}
\end{figure*}

\autoref{fig:RADEX results} shows the RADEX-derived parameters along the blue-shifted lobe. The kinetic temperatures $\Tk$ ranges from 100~K to a few times $10^3$~K. Because the collision data only cover temperatures up to 1000~K, we cannot well constrain the temperature above 1000~K. The SiO column densities $\NSiO$ are in the range of $5\times10^{13}$--$10^{14}$~\sqc{}. The $\rm H_2$ number densities $\nHH$ are on the order of $10^4$--$10^5$~\cc{}. The large uncertainties of the derived physical parameters do not establish a statistically significant axial trend in temperature or density. Assuming the collimated lobe is a cylinder with its axis perpendicular to the line of sight, its line-of-sight depth should be approximately equal to its transverse diameter, $\sim$8000~AU. Adopting this depth and the mean H$_2$ number density of $3\times10^4$~\cc{}, the implied H$_2$ column density is $N\lt({\rm H_2}\rt)=\nHH L\sim10^{21}$~\sqc{}. Combining this with the observed SiO column density $\NSiO$ yields an SiO abundance $X_{\rm SiO}\sim\frac{\NSiO}{N({\rm H_2})}\simeq 2.8\times10^{-8}$.

At the offset of 10\arcsec{}  in \autoref{fig:RADEX results}, which corresponds to SiO \Jff{} enhancement marked in \autoref{fig:Moment0 maps}, we found an elevated molecular number density. Together with the two Class~I \CHthOH{} masers and the linewidth broadening at this position, this position may experience a localized shock interaction. The lobe may interact with dense ambient gas, producing a compressed and kinematically disturbed region \cite[e.g.,][]{bouvier2025}.

\autoref{fig:transverse_T} shows the RADEX-derived kinetic temperatures as well as the integrated intensities of SiO~\Jff{}, \Jes{} transvering the blue-shifted lobe. Because of the asymmetric distribution of the SiO emmission on the two side of the major axis, we are only able to derive temperatures on the northeastern side. The first four temperatures are between $\sim$100 and 500~K while the temperatures near the lobe major axis rise to $\gtrsim$1000~K. The increasing trend of the integrated intensity of SiO~\Jff{} and \Jes{} can also be seen.

\subsubsection{Comparing Temperatures and Column Densities from Different Approaches}\label{subsec:Excitation Conditions of the Outflow}

The kinetic temperatures of several hundred K constrained from RADEX exceed the SiO rotational temperatures, $T_\text{rot}\sim10$--20~K, estimated from rotation diagrams. This difference is consistent with sub-thermal excitation. The SiO column densities from the rotation diagram and forward-modeling method differ. The latter gives values 10 times higher. We attribute this inconsistency to the neglect of optical depth in the rotation diagram method. The derived column density from RADEX analysis is $\sim10^{14}$~\sqc{}, which is consistent with the $10^{13}$--$10^{15}$~\sqc{} range quoted by \citet{Lu2021}. It is $\sim 10^3$ times higher than the nominal LTE values of $10^{10}$--$10^{11}$~\sqc{}. We attribute this to three factors. First, the neglect of optical depth in rotation diagram method can underestimate the column density. Second, high-$J$ SiO emission can be more compact than low-$J$ emission \citep{Hirano2006, Palau2006}. Without fully resolving the compact outflow components, the inclusion of low-$J$ transitions biases the LTE column density toward diffuse gas and thus underestimates the SiO column density in the outflow. 
Third, the sub-thermal conditions ($\Tex \ll \Tk$) cause a steep drop in the contribution from high-$J$ SiO transitions \citep{Herczeg2012}, which leads to an underestimation of the total SiO column density under LTE assumptions.

The potentially varying beam filling factor across the SiO transitions may also lead to an underestimation of the H$_2$ number density. Considering that the high-$J$ SiO transitions typically originate from denser gas \citep{Chandler2000,Hirano2006}, their beam-filling factor $f$ is expected to be smaller than that of the low-$J$ transitions, which usually arise from more diffuse gas. This may lead to a situation where, although the intrinsic intensity of the high-$J$ SiO lines is higher than that of the low-$J$ lines, the observed brightness temperature of the low-$J$ transitions becomes higher due to beam dilution. As such, when we fitted the SiO lines using RADEX and assumed emission fills the beam, we may introduce biases between the estimated physical properties and the true values. Nevertheless, given that the outflow width does not vary dramatically along the collimated component of the lobe, the trends of the measurements along the major axis of the outflow should remain similar to the true trends under the assumption of a uniform beam-filling factor.

\section{Discussion}\label{sec:Disscussion}

\subsection{Asymmetries of the Outflow}
\label{subsec:Asymmetry}

\subsubsection{Different morphology and physical properties between the two lobes}
\label{subsec:asym_twolobes}

As introduced in \autoref{subsec:Spatial Distributions of the Outflow}, the emission of the outflow presents a highly asymmetric morphology between its two lobes. The southeastern blue-shifted lobe is long and narrow, extending to a projected distance of $\sim$0.8~pc, whereas the northwestern red-shifted lobe is shorter, wider, and more diffuse \citep[see also Figure~5 of][]{Lu2021}.

The asymmetry is also found in physical properties of the two lobes. According to \citet{Lu2021}, the SiO~\Jff{} blue-to-red integrated flux and mass ratios are 2.8 and 3.5, respectively. The mass ratios inferred independently from SO, \CHthOH{}, \HtCO, HC$_3$N, and HNCO are between 2.8 and 5.1, with a median of 3.4.  Although the masses have large uncertainties from uncontrained temperature, abundance, and missing fluxes, the common asymmetric feature across six tracers is unlikely to arise from these uncertainies of any one molecule. On the other hand, the two lobes reach comparable maximum ternimal velocities at $\sim$30--34~\kms{} \citep{Lu2021}.  Therefore, we favor an physical or chemical origin in the environment for such lobe-to-lobe asymmetry over difference in their launching process.

Similar asymmetry is seen in infrared emission. \citet{Crowe2025} used the same SiO~\Jff{} data as \citet{Lu2021} and found that the red-shifted lobe aligns with a chain of knots in several NIRCam filters, whereas the entire blue-shifted SiO lobe is invisible in F470N and other NIRCam filters. F470N contains the H$_2$ 0--0 S(9) line, so the non-detection suggests that the blue-shifted lobes lack warm H$_2$ gas. They proposed different extinctions toward the two lobes as a likely explanation.

The different lobe widths may be explained by the turbulent entrainment model of \citet{Li2013}. The dimensionless collimation parameter $\eta$ can be written as
\begin{equation}
 \eta=\frac{p_{\rm cl}}{p_{\rm wind}}
 =\frac{\rho_{\rm cl}\sigma_{\rm cl}^2}{\rho_{\rm wind}v_{\rm wind}^2},
\end{equation}
where $p_{\rm cl}$ and $p_{\rm wind}$ are the pressures of the evelope and the wind for the protostar, respectively, $\rho$ is the volume density, $\sigma_{\rm cl}$ is the velocity dispersion of the envelope, and $v_{\rm wind}$ is the wind velocity. A larger $\eta$ suggests a smaller opening angle. However, this model is not consistent with the observations: it would make the more confined and therefore more collimated lobe harder to propagate, whereas the observed blue-shifted lobe is both narrower and longer. Additional factors such as radial density structure or obstacles along the red-shifted lobe could be the origin of the asymmetry.

Indeed, as shown in \autoref{fig:overview}, there exist several compact hot cores traced by 1.3~mm continuum emission to the northwest of \target{} lying close to the projected continuation of the lobes, which could be the obstacle that hinders the propagation of the red-shifted lobe. The Sgr~C core census gives effective radii of 1030--6560~AU and $n({\rm H_2})=3.3\times10^6$--$2.8\times10^8$~\cc{} assuming a dust temperature of 20~K, while line analyses show that Sgr~C cores contain supersonic motions and are strongly affected by external pressure \citep{Lu2020,Xu2025,Zhang2025}. The resistance of such a core can be equated to the internal pressure and be written as
\begin{equation}
\begin{split}
 p&\sim\rho_{\rm core}\sigma_{\rm core}^2\\
 &=4.7\times10^{-8}
 \left(\frac{n({\rm H_2})}{10^6~{\rm cm}^{-3}}\right)\times
 \left(\frac{\sigma_{\rm core}}{1~{\rm km\,s}^{-1}}\right)^2
 {\rm dyn\,cm}^{-2}.
\end{split}
\end{equation}
Even the lower limit of the density range in the cores gives $p\sim1.5\times10^{-7}$~dyn\,cm$^{-2}$ for a conservative $\sigma_{\rm core}=1$~\kms{}.  For comparison, the RADEX-derived density $n({\rm H_2})\sim5\times10^4$~\cc{} and a characteristic outflow velocity of 7.5--10~\kms{} give an outflow ram pressure $p_{\rm out}=\rho v^2\sim(1.3$--$2.4)\times10^{-7}$~dyn\,cm$^{-2}$. The time for a 10~\kms{} outflow lobe to cross one core diameter is a few $\sim10^3$~yrs. If the projected overlap is physical, encounters can therefore decelerate the red-shifted lobe and produce a shorter, wider, and more fragmented appearance.

Although the true three dimensional positions of the red-shifted lobe and the north-western hot cores are unknown, hints of interaction between them exist. A semi-ring of enhanced SiO emission surrounds the north-western cores and faces the red-shifted lobe (and the JWST/NIRCam F405N emission knots tracing a jet), as shown in \autoref{fig:overview}. It could be a result of impact between the outflow and the dense gas. This will be discussed in a following work focusing on the outflows and shocks in the north-western core group.

\subsubsection{Asymmetric molecular emission perpendicular to the outflow major axis}
\label{subsec:asym_transverse}

A second asymmetry of the outflow is seen in the molecular emission perpendicular to the blue-shifted lobe. Although we visually define the major axis shown in \autoref{fig:overview}, which introduces uncertainties in its exact position, it is clear that the emission of the three molecular lines in \autoref{fig:Moment0 maps} concentrates predominantly on the northeast of it. Nevertheless, the contrast can depend on excitation, shock chemistry, optical depth, and interferometric spatial filtering. Therefore, we refer to it as a transverse asymmetry of the molecular emission, rather than interpreting it as a missing cavity wall or as proof that molecular gas is physically absent.

This one-sided molecular emission distribution may be related to different environmental conditions around the lobe. In particular, the Sgr~C \hii{} region to the western side of the observed field could expand and compress molecular gas, enhancing SiO emission along an interface (see the yellow contours in \autoref{fig:overview}). The hot core itself is embedded in a dense clump \citep{Lu2019a,Lu2024}, and only its southwestern side is exposed to the \hii{} region. This could lead to dissociation, ionization, or reduced shielding and therefore suppress molecular emission on the southwestern side of the blue-shifted lobe. Alternatively, jet precession can change the direction of the outflow and shift shocked gas relative to an axis, leading to an apparent transverse asymmetry \citep{rosen2004}. However, the available data cannot distinguish between these scenarios.

The variation in gas temperature perpendicular to the lobe could be related to feedback from the \hii{} region. As shown in \autoref{fig:transverse_T}, the temperatures near the lobe major axis are higher than those on the northeast. Its far-uv radiation can heat an exposed gas interface, although the exact 3D configuration is unknown and therefore the incident field is unconstrained \citep{Kaufman1999}. The ionization front and magnetic pressure around the \hii{} region can also lead to compression or an oblique shock toward the lobe \citep{Bally2025}. For example, for a typical magnetic field strength of 1~mG, the magnetic pressure is $4\times10^{-8}$~dyn\,\sqc{}. Meanwhile, the thermal pressure across the scale in \autoref{fig:transverse_T} because of the temperature change is
\begin{equation}
\begin{split}
 \Delta p_{\rm th}&\sim n({\rm H_2})k_{\rm B}\Delta T_{\rm k}\\
 &\simeq6.9\times10^{-9}
 \left(\frac{n({\rm H_2})}{5\times10^4~{\rm cm}^{-3}}\right)\left(\frac{\Delta T_{\rm k}}{1000~{\rm K}}\right)
 \ {\rm dyn~cm^{-2}},
\end{split}
\end{equation}
which is one order of magnitude lower than the magnetic pressure. Therefore, the external pressure from the \hii{} region is a feasible but unlikely the only factor causing the asymmetric temperature. 

Finally, the possibility of internal heating cannot be ruled out, since we only derive temperatures on one side of the lobe, using our definition of the major axis. It is possible that stronger shocks in the spine of the outflow heat the gas and produce an increase in temperature toward the spine. However, given the challenge of finding the true outflow major axis, this possibility remains unclear and will not be discussed further.

\subsection{Driving Mechanism of the Outflow from Molecular and Infrared Diagnostics}
\label{subsec:Driving Mechanism of the Outflow}

First, we clarify that the observed molecular outflow traced by SiO is unlikely a jet by itself. The ALMA observations traced shocked SiO and entrained gas on scales of thousands of AU, while the jet or disk wind is usually launched on scales of a few AU or smaller \citep{Bally2016,Moscadelli2022}, which are below the resolution of our observations. Nevertheless, we can infer the launching mechanism of the outflow based on the molecular and infrared observations. We do not aim to resolve the jet-versus-wind dichotomy, because a dense collimated jet could co-exist with a lower-density wide-angle wind \citep[e.g.,][]{arce2013,tabone2017}.

The morphology alone is insufficient to identify the driving mechanism. For example, repeated bow shocks from a narrow jet can sweep up a wide cavity, while wind-driven outflows can become highly collimated under certain conditions \citep{Shang2006,Rabenanahary2022}. In our case, the blue-shifted lobe extends for $\sim$0.8~pc but is only 8000~AU wide over much of its length, giving a projected aspect ratio of about 20. In contrast, the gas closer to the driving core is broader and kinematically more complex. This nested morphology can arise from a fast collimated jet plus a slow disk wind, or from multiple bow-shock wings that excavate a wider cavity close to the hot core \citep[e.g.,][]{raga1993,tafalla2017,lee2022}. The lobe-to-lobe asymmetry discussed in \autoref{subsec:asym_twolobes} also suggests that the morphology is subject to the environment and does not reflect the initial state of the entrained gas. Therefore, the driving mechanism cannot be determined by the morphology.

The PV diagrams provide another diagnostic. In these diagrams, steady jet-driven bow shocks can produce convex spurs, whose highest velocities show up near a bow tip, whereas steady wind-angle winds can produce a more continuous and tilted lobe \citep{Lee2001}. In \autoref{fig:PV diagrams}, the broad emission between $-$65 and $-$45~\kms{} around the hot core is consistent with a wide-angle base of the outflow. However, such emission can also be created by accretion shocks due to infalling gas toward the core \citep{Duarte-Cabral2014}. Between offsets of $\sim$5 and 10\arcsec{}, the lobe becomes narrower near $-$55 to $-$50~\kms{} and its centroid velocity approaches the systemic velocity. This can be produced when a collimated jet transfers momentum to entrained ambient gas and is decelerated \citep[e.g.,][]{codella1999}. Then the linewidth broadens again at offsets of $\sim$10--15\arcsec{}, which is coincident with the increase of molecular number density and the detection of two Class~I \CHthOH{} masers, and therefore could be related to an internal bow shock. In summary, the velocity structure revealed by the PV diagrams is consistent with a jet-driven entrainment.

Next, we consider the excitation conditions along the lobe. The SiO-emitting component is subthermally excited, with $T_\textrm{ex}\sim10$--20~K and effective RADEX kinetic temperatures $T_k$ of several 10$^2$~K. As shown in \autoref{fig:RADEX results}, along the collimated part, $T_k$ ranges between 300 and 800~K and $\log_{10}n({\rm H_2})$ spans $\sim3.3$--4.0, with large uncertainties and no statistically significant monotonic trend. Therefore, we cannot interpret the axial change either as a steady increase of temperature and density or an isothermal profile \citep[e.g.,][]{Liu2018}. Nonetheless, the hot gas component traced by SiO and the presence of Class~I \CHthOH{} masers suggests active shock processing inside the lobe, which can be generated by a jet \citep{Lee2001}. In addition, in \autoref{subsec:asym_transverse} we show that external compression could contribute to the temperature distribution in the lobe. Therefore, excitation in the lobe alone cannot distinguish the driving mechanism.

Lastly, we examine the near-infrared emission from the JWST/NIRCam data \citep{Crowe2025}, and find evidence that the red-shifted lobe in the northwest is driven by a jet. \citet{Crowe2025} identified G359.44b as the massive protostar driving the outflow and found that the red-shifted SiO emission aligns with the infrared knots K34, K32, and K27 (see \autoref{fig:overview}). These knots are detected in multiple filters, including F470N, which covers H$_2$ 0--0 S(9), and F405N, which covers Br$\alpha$. It is interesting to note that the entire blue-shifted lobe is invisible in F470N and other NIRCam filters. As discussed in \autoref{subsec:asym_twolobes}, it can be attributed to differential extinction. Therefore, the lack of an infrared counterpart does not rule out the existence of a blue-shifted jet aligning with the outflow lobe in the southeast.

Taken together, the \target{} outflow is likely primarily driven by a jet, while the broad component close to the hot core could be related to a wide-angle wind. Future higher-resolution observations toward the base of the outflow have the potential to directly resolve the proposed jet.

\subsection{Implications to High-mass Star Formation in the Central hot core and the CMZ}

We first estimate the outflow mass. Using the column density derived from RADEX in \autoref{subsubsec:Non-Local Thermodynamic Equilibrium Method}, and assuming a beam filling factor of 1, we derive the outflow mass
\begin{equation}
 M_{\rm out}=\frac{\mu_{\rm H_2}m_{\rm H}}{X({\rm SiO})}
 \int N({\rm SiO})\,dA,
\end{equation}
where X({\rm SiO}) is the SiO abundance and $\mu_{\rm H_2}=2.8$ is the gas mass per H$_2$ molecule in units of $m_{\rm H}$. Integrating the column density map within the area where all three SiO emission intensities are above $1\times$ RMS noise levels gives $\sim5.7$~\msun{} for $\NSiO=10^{14}$~\sqc{} and $X({\rm SiO})=2.8\times10^{-8}$. 
The uncertainty in the estimated mass is large, given the errors in, e.g., temperature, optical depth, and abundance. \citet{Lu2021} estimated a typical factor of $\sim$70 uncertainty for the mass of a CMZ outflow sample that includes \target{}.

The momentum $p_{\rm out}$ and the mechanical force $F_{\rm out}$ are measured relative to the source velocity $v_\textrm{sys}=-50.2$~\kms{},
\begin{equation}
 p_{\rm out}=\int |v-v_{\rm sys}|\,dM(v),
\end{equation}
\begin{equation}
 F_{\rm out}=\frac{p_{\rm out}}{t_{\rm dyn}}.
\end{equation}
However, the channel-dependent mass $M(v)$ is not available because we use integrated intensities for the RADEX analysis. Using the SiO blue-lobe mass of 2.4~\msun{} and its terminal velocity offset of 10~\kms{}, we estimate $p_{\rm blue}\lesssim147$~\msun{}~km~s$^{-1}$. The true momentum should be smaller, because most of the mass is not at the terminal velocity. With a projected dynamical time of $3.2\times10^4$~yr, we estimate $F_{\rm blue}\lesssim4.6\times10^{-3}$~\msun{}~km~s$^{-1}$~yr$^{-1}$ before inclination correction.

For a magnetocentrifugal wind, the wind momentum injection rate can be written as $\dot p_{\rm w}=\phi_{\rm w}\dot M_*v_K$, with $\dot M_*$ the accretion rate of the protostar, $v_K$ the Keplerian velocity, and $\phi_{\rm w}\sim0.2$ for a representative disk wind \citep{Tan2014}. Assuming a fraction $\epsilon_p$ of the momentum is transferred to the observed outflow, the accretion rate can be written as
\begin{equation}
 \dot M_*\sim{}\frac{2.4\times10^{-4}}{\epsilon_p}
 \left(\frac{F_{\rm out}}{F_0}\right)
 \left(\frac{0.2}{\phi_{\rm w}}\right)\left(\frac{100~{\rm km~s^{-1}}}{v_K}\right)\ M_\odot\,{\rm yr}^{-1}.
\end{equation}
Here the normalization factor $F_0$ is $4.6\times10^{-3}$~\msun{}\,\kms{}\,yr$^{-1}$, while $\epsilon_p$ and $v_K$ cannot be constrained from the observations. For comparison, \citet{Crowe2025} used radiative-transfer modeling of the spectral energy distribution to estimate an accretion rate of $\sim$$4.2\times10^{-4}$~\msun{}\,yr$^{-1}$ for G359.44b. The two results agree at the order-of-magnitude level, which supports active ongoing accretion toward G359.44b.

With such a high accretion rate, the stellar mass in the core can be substantially increased. The SED fitting in \citet{Crowe2025} gives $M_*\simeq20.4^{+24.1}_{-11.0}$~\msun{}. The corresponding growth time $M_*/\dot M_*\sim8.5\times10^4$~yr is very close to the stellar age from the SED fitting ($\sim$$8.6\times10^4$~yr) as well as the dynamical time of the outflow. If the accretion rate persists for another $10^4$~yr, $\sim$2.4~\msun{} would be transferred through the disk onto the protostar. This suggests that disk-mediated accretion and ejection operate around a massive protostar despite the high-pressure and turbulent environment in the CMZ \citep[see e.g.,][]{Lu2022,Weng2026}. In addition, the analysis in \autoref{subsec:Asymmetry} indicates that the neighboring \hii{} region could reshape the pc-scale entrained gas, although the inner accretion flow could still be intact \citep{Tan2014,Kuiper2015}.

Within the sample of 43 CMZ outflows in \citet{Lu2021}, \target{} occupies the high-mass and large-size end. Outflow masses in this sample range from $\lesssim$1~\msun{} to a few tens of solar masses, and the 0.8-pc lobe of \target{} is among the longest. \target{} is much more massive than the outflows in G0.253$+$0.016, where nine of 18 dense cores are associated with bipolar SiO outflows but no high-mass protostar was identified \citep{Walker2021}. Their outflow masses, scaled to the SiO abundance of $10^{-9}$ used here, are all $\lesssim$5~\msun{}.
At the other extreme, the Sgr~B2(N) K2 outflow is reported to have a mass of $\sim$2000~\msun{} and an opening angle of $\sim$60\arcdeg{} \citep{Higuchi2015}. The \target{} outflow bridges the compact lower-mass outflows and the massive and less collimated outflows in Sgr~B2. These detections of outflows together suggest that the disk-jet picture in high-mass star formation remains valid in the CMZ, while the environment can modify the large-scale propogation of the entrained lobes.

\section{Summary}\label{sec:conclusion}

We have combined ALMA observations of three SiO transitions with archival JWST images to resolve the morphology, kinematics, and excitation of the highly collimated \target{} outflow. Our main findings are:
\begin{itemize}
\item The blue-shifted lobe in the southeast extends for $\gtrsim$0.8~pc while remaining only about 8000~AU wide, giving a projected aspect ratio of 20. A broad and kinematically complex component is found near the driving hot core, whereas three Class~I \CHthOH{} masers and localized SiO enhancements lie along the narrow lobe. The red-shifted lobe in the northwestern is shorter and wider.
\item SiO rotational temperatures of 10--20~K are lower than the non-LTE kinetic temperatures of several hundred K, suggesting sub-thermal excitation. Column densities and H$_2$ number densities are model dependent. 
\item The asymmetry seen in molecular and infrared emission between the two lobes and perpendicular to the blue-shifted lobe is compatible with uneven environmental impacts and extinction. Dense cores along the projected path of the red-shifted lobe is a possible obstacle, but a physical interaction is not established. Transversing the detected blue-shifted gas, the kinetic temperature rises toward the visually adopted spine, which may be related to feedback from the nearby Sgr~C \hii{} region.
\item We suggest an important role for jet-driven entrainment, while a broader wind is not excluded. The association with the massive protostar G359.44b supports continuing accretion-ejection activity in the CMZ, similar to high-mass star formation elsewhere in the Galaxy.
\end{itemize}

\begin{acknowledgements}
We thank Prof.\ Keping Qiu, and Drs.\ Suinan Zhang and Yu Cheng for helpful discussions. This work has been supported by the Strategic Priority Research Program of the Chinese Academy of Sciences (CAS) Grant No.\ XDB0800300, the National SKA Program of China (2025SKA0140100), the National Natural Science Foundation of China (NSFC) through grant Nos.\ 12273090 and 12322305, and the National Key R\&D Program of China (No.\ 2022YFA1603101). This paper makes use of the following ALMA data: ADS/JAO.ALMA\#2016.1.00243.S, 2018.1.01420.S, 2021.1.00286.S, and 2024.1.00213.S. ALMA is a partnership of ESO (representing its member states), NSF (USA) and NINS (Japan), together with NRC (Canada), MOST and ASIAA (Taiwan), and KASI (Republic of Korea), in cooperation with the Republic of Chile. The Joint ALMA Observatory is operated by ESO, AUI/NRAO and NAOJ\@. This research has made use of NASA’s Astrophysics Data System.
\end{acknowledgements}

\vspace{5mm}
\facilities{ALMA, JWST, MeerKAT}
\software{CASA \citep{CASA2022}, CARTA \citep{carta2026}, radio-beam \citep{koch2025}, \texttt{emcee} \citep{emcee2013}}

\bibliography{references} 
\bibliographystyle{aasjournalv7}

\end{CJK}
\end{document}

%% file: authors.tex
\author[0009-0009-5591-9042]{Sen-di Bo}
\affiliation{Shanghai Astronomical Observatory, Chinese Academy of Sciences, 80 Nandan Road, Shanghai 200030, P.\ R.\ China}
\affiliation{School of Astronomy and Space Sciences, University of Chinese Academy of Sciences, No. 19A Yuquan Road, Beijing 100049, P.\ R.\ China}
\email{bosendi26@mails.ucas.ac.cn}

\author[0000-0003-2619-9305]{Xing Lu}
\affiliation{State Key Laboratory of Radio Astronomy and Technology, Shanghai Astronomical Observatory, Chinese Academy of Sciences, 80 Nandan Road, Shanghai 200030, P.\ R.\ China}
\email{xinglu@shao.ac.cn}

\author[0000-0002-9776-5610]{Tian-Yu Tu}
\affiliation{School of Astronomy \& Space Science, Nanjing University, 163 Xianlin Avenue, Nanjing 210023, P.\ R.\ China}
\affiliation{Laboratoire d'Astrophysique de Bordeaux, Univ.\ Bordeaux, CNRS, B18N, all\'{e}e Geoffroy Saint-Hilaire, 33615 Pessac, France}
\email{tianyu.tu1105@outlook.com}

\author[]{Weiyuan Zhang}
\affiliation{Shanghai Pinghe School, 333 Shenqi Road, Pudong District, Shanghai, P.\ R.\ China}
\email{vivianzhang0956@163.com}

\author [0009-0005-0394-3754] {Samuel Crowe}
\affiliation{Oxford Centre for the History of Science, Medicine, and Technology, University of Oxford, Oxford OX2 6AH, United Kingdom}
\email{scroweastro@gmail.com}

\author[0000-0002-4774-2998]{Junhao Liu}
\affiliation{School of Astronomy and Space Science, Nanjing University, 163 Xianlin Avenue, Nanjing, Jiangsu 210023, P.\ R.\ China}
\affiliation{Key Laboratory of Modern Astronomy and Astrophysics (Nanjing University), Ministry of Education, Nanjing, Jiangsu 210023, P.\ R.\ China}
\email{liujunhao@nju.edu.cn}

\author[0000-0003-1275-5251]{Shanghuo Li}
\affiliation{School of Astronomy and Space Science, Nanjing University, 163 Xianlin Avenue, Nanjing, Jiangsu 210023, P.\ R.\ China}
\affiliation{Key Laboratory of Modern Astronomy and Astrophysics (Nanjing University), Ministry of Education, Nanjing, Jiangsu 210023, P.\ R.\ China}
\email{shli@nju.edu.cn}

\author[0000-0001-8135-6612]{John Bally}
\affiliation{Center for Astrophysics and Space Astronomy, Department of Astrophysical and Planetary Sciences University of Colorado, Boulder, CO 80389, USA}
\email{john.bally@colorado.edu}